\documentclass[10pt]{article}
\usepackage[latin1]{inputenc}
\usepackage[english]{babel}
\usepackage[none]{hyphenat}
\usepackage[T1]{fontenc}
\usepackage[left=2cm,right=2cm,top=1.5cm,bottom=2cm]{geometry}
\usepackage[usenames,dvipsnames]{color}  
\usepackage[pdfborder={0 0 0},colorlinks=true,urlcolor=blue,linkcolor=blue,citecolor=blue,bookmarks=true]{hyperref}

\hypersetup{pdftitle={Bandicoot: C++ Library for GPU Linear Algebra and Scientific Computing}}
\hypersetup{pdfauthor={Ryan R. Curtin, Marcus Edel, Conrad Sanderson}}

\usepackage{newtxtext}  
\usepackage{inconsolata}   

\usepackage{microtype}
\DisableLigatures[f]{encoding = *, family = * }

\usepackage{titlesec}
\usepackage{graphicx}
\usepackage[frozencache=true,cachedir=minted-cache]{minted}
\usepackage{amsmath}
\usepackage{amsthm}
\usepackage{bm}
\usepackage{booktabs}      
\usepackage{multirow}      
\usepackage{tcolorbox}   
\usepackage{adjustbox}
\usepackage{threeparttable}
\usepackage{parcolumns}
\usepackage{amsfonts}
\usepackage{nicefrac}
\usepackage{microtype}
\usepackage{tabularx}
\usepackage{array}
\usepackage{wasysym}
\usepackage{subcaption}

\titleformat*{\section}{\bf\normalsize\large}
\titleformat*{\subsection}{\bf\normalsize}

\titlespacing{\section}{0pt}{1.5ex}{0ex}
\titlespacing{\subsection}{0pt}{1.5ex}{0ex}
\titlespacing{\subsubsection}{0pt}{1.5ex}{0ex}

\usepackage{microtype}
\DisableLigatures[f]{encoding = *, family = * }

\small\normalsize

\newcolumntype{R}[2]{%
  >{\adjustbox{angle=#1,lap=\width-(#2)}\bgroup}%
  l%
  <{\egroup}%
}

\begin{document}

\pagestyle{empty}

\begin{center}


\scalebox{1.5}{\bf Bandicoot: C++ Library for GPU Linear Algebra and Scientific Computing}

\vspace{1.5ex}
Ryan R. Curtin{\tiny~}{$^{1}$},
Marcus Edel{\tiny~}{$^{2,3}$}, 
Conrad Sanderson{\tiny~}{$^{4,5}$}
\vspace{1.5ex}


\begin{small}
\begin{tabular}{l}
$^{1}$~Independent Researcher, Atlanta, USA\\
$^{2}$~Free University of Berlin, Germany\\
$^{3}$~Collabora, Montreal, Canada\\
$^{4}$~Data61~/~CSIRO, Australia\\
$^{5}$~Griffith University, Australia\\
\end{tabular}
\end{small}

\end{center}

\section*{Abstract}

This report provides an introduction to the {\bf Bandicoot} C++ library for
linear algebra and scientific computing on GPUs,
overviewing its user interface and performance characteristics,
as well as the technical details of its internal design.
Bandicoot is the GPU-enabled counterpart to the well-known Armadillo C++ linear
algebra library,
aiming to allow users to take advantage of GPU-accelerated computation
for their existing codebases without significant changes.
Adapting the same internal template meta-programming techniques that Armadillo uses,
Bandicoot is able to provide compile-time optimisation of mathematical expressions within user code.
The library is ready for production use
and is available at \url{https://coot.sourceforge.io}.
Bandicoot is distributed under the Apache 2.0 License.

\vspace{4ex}
\hrule

\section{Introduction}
\label{sec:introduction}

A number of recent significant advancements in scientific computing have been
enabled by the use of graphics accelerators / graphics processing units (GPUs),
notably in the fields of machine learning~\cite{ciresan2011flexible, krizhevsky2012imagenet},
HPC and supercomputing~\cite{hamada200942, stone2013gpu},
healthcare and life sciences~\cite{ravi2016deep, vamathevan2019applications},
as well as modeling and simulation~\cite{harris2003simulation}.
GPUs are built on the concept of SIMT (single instruction, multiple threads)
and are thus capable of performing the same operation on separate chunks of data in parallel.
Often, modern GPUs are extremely parallel:
Nvidia's recent mainstream consumer-focused GeForce RTX~4090 GPU contains 16,384 cores,
leading to a throughput of around 80 TFLOPS for 32-bit floating point data.

A large part of the reasons for the significant speedups and advancements in the fields listed above
is the core accelerations that GPUs provide to linear algebra operations.
The nature of most linear algebra computations is very well-suited to SIMT-like approaches:
for instance, the common {\tt axpy} vector-scalar product task,
(e.g. $\mathbf{y} \gets \mathbf{y} + \alpha \mathbf{x}$)
is embarrassingly parallel and clearly suited for SIMT processing on a GPU.

However, from a user perspective,
writing scientific code that properly exploits GPUs is cumbersome and not portable.
The CUDA framework~\cite{nickolls2008scalable} is specific to Nvidia devices,
and thus any CUDA code cannot directly work on AMD or Intel GPUs,
or any other GPU-like devices such as Google's tensor processing units~\cite{jouppi2017datacenter}.
Even higher-level toolkits such as MATLAB~\cite{MATLAB} and R~\cite{Rmanual}
typically need special setup to use GPUs,
and that setup is typically limited to a certain type of GPU.
Standardization efforts like OpenCL~\cite{stone2010opencl},
which promise code that works on any GPU-like device,
are often not well-supported by vendors.
Balkanization has occurred in the space,
with numerous incompatible efforts like
Apple's Metal\footnote{\url{https://developer.apple.com/metal/}} and
AMD's ROCm\footnote{\url{https://www.amd.com/en/graphics/servers-solutions-rocm}},
in addition to Nvidia's CUDA.

From a non-expert user's perspective,
this situation is abysmal:
scientists and developers operating at a higher level do not want to spend time digging into low-level implementation details of their GPUs,
but instead want to be able to simply write linear algebra and have it work efficiently on their GPU.
This is, after all, the reason that linear algebra libraries and interfaces like LAPACK and BLAS~\cite{anderson1999lapack, blackford2002updated} were originally written:
to provide an interface in order to abstract away the implementation details.

High-level interfaces have seen massive amounts of success over the years,
specifically because they allow developers and scientists to maximize productivity by focusing on their higher-level task.
In this vein, toolkits like MATLAB and Mathematica~\cite{wolfram1999mathematica} have been extremely successful,
in large part because of the simple and intuitive interface they provide.
In the field of machine learning, Python~\cite{Python3_manual} 
and its linear algebra abstraction NumPy~\cite{harris2020array} have become dominant in large part for the same reason.
However, these higher-level languages often come with a cost;
for instance, MATLAB and Python have high amounts of overhead,
and require large runtime environments that depend on numerous additional packages.

Motivated by high-level MATLAB-style expressive syntax,
the well-known Armadillo C++ linear algebra library~\cite{sanderson2016armadillo} was created
to satisfy the need for a high-level and intuitive linear algebra interface
while maximizing runtime performance.
Armadillo uses template meta-programming~\cite{Vandevoorde_2017} in order to optimise linear algebra expressions at program compilation time;
this can provide significant and demonstrable speedups over other
solutions~\cite{sanderson2018user, mca24030070, sanderson2020adaptive, sanderson2017armadillo}.
Its clear interface and flexibility has led to widespread adoption,
including via bindings to other languages like R~\cite{eddelbuettel2014rcpparmadillo} and Python~\cite{Urlus_CARMA_2023,rumengan2021pyarmadillo}.

However, Armadillo is only able to use CPUs and CPU-specific libraries
such as OpenBLAS~\cite{OpenBLAS_2023,wang2013augem} and Intel's Math Kernel Library (MKL) for its linear algebra operations.
Therefore, practitioners cannot make use of the GPUs that they likely have available to accelerate their linear algebra programs.
This motivated us to develop the {\bf Bandicoot} C++ GPU linear algebra library.

Bandicoot is aiming to provide an API that closely matches Armadillo's API,
with the primary difference being that linear algebra operations are performed on the GPU instead of the CPU,
and the memory used for matrices is GPU memory instead of system RAM.
Bandicoot abstracts away the vendor-specific technologies that are used for GPU programming,
and can run on virtually any GPU via its support for both CUDA and OpenCL backends.
Users do not need to learn low-level details about their devices,
as Bandicoot will automatically try to configure and use the most appropriate device.

Higher-level libraries are already beginning to use Bandicoot,
with work underway to adapt the {\it ensmallen} numerical optimisation library~\cite{curtin2021ensmallen}
and the {\it mlpack} machine learning library~\cite{curtin2023mlpack}
to use Bandicoot and provide GPU-accelerated algorithms.

The rest of this report is dedicated to
describing the user interface of Bandicoot (Section~\ref{sec:overview}),
detailing differences from Armadillo that are important to know when converting Armadillo code to use GPUs (Sections~\ref{sec:arma_comparison} and~\ref{sec:linking}),
demonstrating speedups gained by using Bandicoot (Section~\ref{sec:speed}),
defining the upcoming roadmap of features to be implemented (Section~\ref{sec:roadmap}),
and documenting the internal design and details of the library (Section~\ref{sec:internals}).

\section{Overview of Functionality}
\label{sec:overview}

Bandicoot aims to provide an API that closely matches the API of the Armadillo C++ linear algebra library~\cite{sanderson2016armadillo}.
This means that Bandicoot's interface is the same as Armadillo's intuitive and readable interface,
so far as is possible (see Section~\ref{sec:arma_comparison}).
The following example Bandicoot program may therefore look familiar to Armadillo users:

{\footnotesize
\begin{minted}{c++}
    #include <bandicoot>

    using namespace coot;

    int main()
      {
      fmat X(1000, 1000, fill::randu);
      fmat Y(1000, 2000, fill::randu);

      // (X + I)' (Y + 2)
      fmat Z = (X + eye<fmat>(1000, 1000)).t() * (Y + 2);
      }
\end{minted}
}

In fact, due to Bandicoot's API-compatibility,
the code above can be easily changed to be valid Armadillo code;
one simply needs to replace ``{\tt \#include <bandicoot>}'' with ``{\tt \#include <armadillo>}'',
and ``{\tt using namespace coot;}'' with ``{\tt using namespace arma;}''.

The basic type in Bandicoot is the {\tt Mat} (matrix) class,
which as a template parameter takes the element type to be held in the matrix.
That is, the {\tt Mat<float>} class is a matrix that holds 32-bit {\tt float} values.
Bandicoot provides convenient {\tt typedef}s for various matrix types;
thus, the {\tt fmat} class is equivalent to {\tt Mat<float>},
and {\tt umat} class is equivalent to {\tt Mat<uword>},
where {\tt uword} is the unsigned integer type provided by Bandicoot---generally
equivalent to {\tt size\_t}.

Similar to the {\tt Mat} class,
Bandicoot provides the {\tt Col} and {\tt Row} classes for row and column vectors, respectively.
Convenience {\tt typedef}s are also provided:
{\tt fvec} and {\tt fcolvec} for {\tt Col<float>},
and {\tt frowvec} for {\tt Row<float>}.

When Bandicoot code is run,
linear algebra operations are performed on the GPU.
Bandicoot supports both a CUDA backend and an OpenCL backend.
Details of which backend to be used and which device to use
can optionally be controlled via the {\tt coot\_init()} function,
meant to be used at the beginning of the program:

{\footnotesize
\begin{minted}{c++}
// Use only one of the following in your program, or omit entirely for automatic selection
coot_init("opencl", true /* print information */);
coot_init("opencl", false, 0 /* OpenCL platform ID */, 1 /* OpenCL device id */);
coot_init("cuda", true, 2 /* CUDA device ID */);
\end{minted}
}

Of course, device initialisation will fail if neither OpenCL nor CUDA is available on the system.

More complicated linear algebra operations,
such as matrix multiplication and decompositions (LU, SVD, eigen, etc.),
use external support libraries whenever possible.
The support libraries that are used depend on which backend is enabled:
\vspace*{-1em}
\begin{itemize}
  \itemsep -2pt
  \item The {\bf CUDA} backend uses cuBLAS, cuRand, and cuSolver, which are parts of the CUDA Toolkit\footnote{\url{https://developer.nvidia.com/cuda-toolkit}}.
  \item The {\bf OpenCL} backend uses clBLAS~\cite{clblas} and internal adaptations of MAGMA~\cite{tdb10} and clMAGMA~\cite{hcydglt14}.
\end{itemize}
\vspace*{-1em}
This is similar to Armadillo's use of external support libraries
such as OpenBLAS~\cite{wang2013augem} or other BLAS/LAPACK replacements.

A condensed overview of the functionality%
of Bandicoot is given in Tables~\ref{tab:mat_members} through~\ref{tab:func_matrices}:

\vspace*{-1em}
\begin{itemize}
\itemsep -2pt
\item Table~\ref{tab:mat_members} briefly describes the main members of the {\tt Mat}
class.
\item Table~\ref{tab:operators} lists the overloaded C++ operators that can be used with Bandicoot objects.
\item Table~\ref{tab:decompositions} lists a subset of currently available decomposition techniques.
\item Table~\ref{tab:generators} lists a subset of functions that generate matrices and vectors.
\item Table~\ref{tab:func_scalars} lists a subset of functions for Bandicoot objects that return scalar values.
\item Table~\ref{tab:func_vectors} lists a subset of functions for Bandicoot objects that return vectors ({\tt Col}/{\tt Row}).
\item Table~\ref{tab:func_matrices} lists a subset of functions for Bandicoot objects that return matrices ({\tt Mat}).
\end{itemize}

The full documentation is available online at \url{https://coot.sourceforge.io/docs.html}.

\begin{table}[h!]
\begin{center}
\begin{tabular}{l|l}
\hline
{\bf Function/Variable} & {\bf Description} \\
\hline
{\tt .n\_rows} & number of rows (read-only) \\
{\tt .n\_cols} & number of columns (read-only) \\
{\tt .n\_elem} & total number of elements (read-only) \\
\hline
 & {{\bf Note:} individual element access is inefficient; see Section~\ref{sec:arma_comparison}.} \\
{\tt (i)} & access the {\tt i}-th element, assuming a column-by-column layout \\
{\tt [i]} & access the {\tt i}-th element with no bounds check \\
{\tt (r, c)} & access the element at row {\tt r} and column {\tt c} \\
{\tt .at(r, c)} & access the element at row {\tt r} and column {\tt c}, but with no bounds check \\
{\tt .get\_dev\_mem()} & get raw GPU memory (returns object with {\tt .cuda\_mem\_ptr}/{\tt .cl\_mem\_ptr} members) \\
\hline
{\tt .reset()} & set the number of elements to zero \\
{\tt .set\_size(rows, cols)} & set the size to the specified dimensions, without preserving data \\
{\tt .reshape(rows, cols)} & set the size to the specified dimensions, while preserving data \\
\hline
{\tt .fill(k)} & set all elements to be equal to {\tt k} \\
{\tt .ones(rows, cols)} & set all elements to one, optionally first resizing to specified dimensions \\
{\tt .zeros(rows, cols)} & set all elements to zero, optionally first resizing to specified dimensions \\
{\tt .eye(rows, cols)} & set to the identity matrix, optionally first resizing to specified dimensions \\
{\tt .randu(rows, cols)} & set to uniformly distributed random values in $[0, 1]$ with optional resize \\
{\tt .randn(rows, cols)} & set to Gaussian random values ($\sim N(0, 1)$) with optional resize \\
\hline
{\tt .t()} & return transpose of object \\
\hline
{\tt .is\_empty()} & test whether there are no elements \\
{\tt .is\_square()} & test whether the matrix is square \\
{\tt .is\_vec()} & test whether object is a vector \\
\hline
{\tt .print(header)} & print elements to the {\tt cout} stream, with an optional header \\
{\tt .raw\_print(header)} & as above, but with no formatting of the output stream \\
\hline
{\it submatrix views:} & \\
\ {\tt .diag(k)} & read/write access to the {\tt k}-th diagonal \\
\ {\tt .row(i)} & read/write access to the {\tt i}-th row \\
\ {\tt .col(j)} & read/write access to the {\tt j}-th column \\
\ {\tt .rows(a, b)} & read/write access to the submatrix spanning from row {\tt a} to row {\tt b} \\
\ {\tt .cols(c, d)} & read/write access to the submatrix spanning from column {\tt c} to column {\tt d} \\
\ {\tt .submat(p, q, r, s)} & read/write access to the submatrix starting at element {\tt (p, q)} and ending at {\tt (r, s)} \\
\ {\tt .submat(span(p, r), span(q, s))} & same as above \\
\hline
\end{tabular}
\end{center}
\caption{Member functions and variables of the {\tt Mat} class (subset).}
\label{tab:mat_members}
\end{table}

\begin{table}[h!]
\begin{center}
\begin{tabular}{l|l}
\hline
{\bf Operation} & {\bf Description} \\
\hline
{\tt A - k} & subtract scalar {\tt k} from all elements in {\tt A} \\
{\tt k - A} & subtract each element in {\tt A} from scalar {\tt k} \\
{\tt A + k}, {\tt k + A} & add scalar {\tt k} to all elements in {\tt A} \\
{\tt A * k}, {\tt k * A} & multiply matrix {\tt A} by scalar {\tt k} \\
\hline
{\tt A + B} & add matrices {\tt A} and {\tt B} \\
{\tt A - B} & subtract matrix {\tt B} from {\tt A} \\
{\tt A * B} & matrix multiplication of {\tt A} and {\tt B} \\
{\tt A \% B} & element-wise multiplication of {\tt A} and {\tt B} \\
{\tt A / B} & element-wise division of {\tt A} by {\tt B} \\
\hline
\end{tabular}
\end{center}
\caption{Matrix operations involving overloaded C++ operator functions (subset).}
\label{tab:operators}
\end{table}

\begin{table}[h!]
\begin{center}
\begin{tabular}{l|l}
\hline
{\bf Function} & {\bf Description} \\
\hline
{\tt lu(L, U, P, X)} & Lower-upper decomposition of {\tt X} such that {\tt P * X = L * U} and {\tt X = P.t() * L * U} \\
{\tt lu(L, U, X)} & Same as above, but with permutation matrix combined with {\tt L}: {\tt X = L * U} \\
{\tt chol(X)} & Cholesky decomposition of {\tt X}, returns {\tt R} such that {\tt X = R.t() * R} \\
{\tt svd(X)} & Singular value decomposition of {\tt X} \\
{\tt eig\_sym(X)} & Eigendecomposition of {\tt X} \\
{\tt pinv(X)} & Moore-Penrose pseudo-inverse of {\tt X} \\
\hline
{\tt solve(A, B)} & Solve a system of linear equations {\tt A * X = B}, where {\tt X} is unknown \\
\hline
\end{tabular}
\end{center}
\caption{Matrix decompositions and related functions (subset).}
\label{tab:decompositions}
\end{table}

\begin{table}[h!]
\begin{center}
\begin{tabular}{l|l}
\hline
{\bf Function} & {\bf Description} \\
\hline
{\tt eye(rows, cols)} & matrix with the elements along the main diagonal set to one; \\
 & if {\tt rows == cols}, an identity matrix is generated \\
\hline
{\tt ones(rows, cols)} & matrix with all elements set to one \\
{\tt zeros(rows, cols)} & matrix with all elements set to zero \\
\hline
{\tt randu(rows, cols)} & matrix with uniformly distributed random values in the interval $[0, 1]$ \\
{\tt randn(rows, cols)} & matrix with random values from the normal distribution with $\mu = 0$ and $\sigma = 1$ \\
\hline
{\tt linspace(start, end, n)} & vector with {\tt n} elements, linearly increasing from {\tt start} up to (and including) {\tt end} \\
{\tt repmat(A, p, q)} & replicate matrix {\tt A} in a block-like fashion, resulting in {\tt p} by {\tt q} blocks of {\tt A} \\
\hline
\end{tabular}
\end{center}
\caption{Functions for generating matrices (subset).}
\label{tab:generators}
\end{table}

\begin{table}[h!]
\begin{center}
\begin{tabular}{l|l}
\hline
{\bf Function} & {\bf Description} \\
\hline
{\tt accu(A)} & accumulate (sum) all elements of {\tt A} \\
{\tt as\_scalar(expr)} & evaluate an expression that results in a $1 \times 1$ matrix, convert the result to a scalar \\
{\tt det(A)} & determinant of square matrix {\tt A} \\
{\tt dot(A, B)} & dot product of {\tt A} and {\tt B}, assuming they are vectors with equal numbers of elements \\
{\tt norm(A, p)} & {\tt p}-norm of {\tt A}, with {\tt p = 1, 2,} $\cdots$ or {\tt p = "inf", "-inf", "fro"} \\
{\tt trace(A)} & sum of the diagonal elements of square matrix {\tt A} \\
\hline
{\tt all(A)} & check whether all elements in {\tt A} are nonzero \\
{\tt all(A < x)} & check whether all elements in {\tt A} are less than scalar {\tt x} \\
 & (note that other relational expressions may also be used) \\
{\tt any(A)} & check whether any elements in {\tt A} are nonzero \\
{\tt any(A < x)} & check whether any elements in {\tt A} are less than scalar {\tt x} \\
 & (note that other relational expressions may also be used) \\
\hline
\end{tabular}
\end{center}
\caption{Scalar valued functions (subset).}
\label{tab:func_scalars}
\end{table}

\begin{table}[h!]
\begin{center}
\begin{tabular}{l|l}
\hline
{\bf Function} & {\bf Description} \\
\hline
{\tt max(A, dim)} & find the maximum in each column of {\tt A} ({\tt dim = 0}), \\
 & or each row of {\tt A} ({\tt dim = 1}) (default: {\tt dim = 0}) \\
{\tt min(A, dim)} & as above, but find the minimum \\
{\tt sum(A, dim)} & as above, but find the sum \\
\hline
{\it statistics:} \\
\ {\tt mean(A, dim)} & as above, but find the average \\
\ {\tt median(A, dim)} & as above, but find the median \\
\ {\tt stddev(A, dim)} & as above, but find the standard deviation \\
\ {\tt var(A, dim)} & as above, but find the variance \\
\hline
{\tt diagvec(A, k)} & extract the {\tt k}-th diagonal from matrix {\tt A} (default: {\tt k = 0}) \\
\hline
\end{tabular}
\end{center}
\caption{Vector valued functions (subset).}
\label{tab:func_vectors}
\end{table}

\begin{table}[h!]
\begin{center}
\begin{tabular}{l|l}
\hline
{\bf Function} & {\bf Description} \\
\hline
{\tt cor(A, B)} & the element at {\tt (i, j)} in the result will be the the correlation \\
 & between the {\tt i}-th variable in {\tt A} and the {\tt j}-th variable in {\tt B}, \\
 & where each row of {\tt A} and {\tt B} is an observation and each column is a variable \\
{\tt cov(A, B)} & as per {\tt cor(A, B)}, but calculate the covariance \\
\hline
{\tt conv(A, B)} & one-dimensional convolution of {\tt A} and {\tt B}, assuming they are vectors \\
{\tt conv2(A, B)} & two-dimensional convolution of matrices {\tt A} and {\tt B} \\
{\tt cross(A, B)} & cross product of {\tt A} and {\tt B}, assuming they are 3-dimensional vectors \\
\hline
{\tt find(A)} & find the indices of non-zero elements of {\tt A} \\
{\tt find(A < k)} & find the indices of elements of {\tt A} with values less than scalar {\tt k} \\
 & (note that other relational expressions may also be used) \\
{\tt find\_finite(A)} & find the indices of finite elements of {\tt A} \\
{\tt find\_nonfinite(A)} & find the indices of nonfinite elements of {\tt A} \\
{\tt find\_nan(A)} & find the indices of elements of {\tt A} with value {\tt nan} (not a number) \\
\hline
{\tt join\_rows(A, B)} & append each row of {\tt B} to its respective row of {\tt A} \\
{\tt join\_cols(A, B)} & append each column of {\tt B} to its respective column of {\tt A} \\
{\tt reshape(A, r, c)} & copy {\tt A} with dimensions set to {\tt r} rows and {\tt c} columns \\
\hline
{\tt sort(A, dim)} & copy {\tt A} with the elements sorted in each column ({\tt dim = 0}) \\
 & or row ({\tt dim = 1}) \\
{\tt sort\_index(A)} & assuming {\tt A} is a vector, generate a vector that describes the \\
 & sorted order of {\tt A}'s elements (i.e. the indices) \\
\hline
{\tt trans(A)} & Hermitian transpose of {\tt A} \\
\hline
{\tt diagmat(A)} & interpret matrix {\tt A} as a diagonal matrix \\
 & (this can save computation time during multiplication) \\
\hline
{\tt misc(A)} & apply a miscellaneous function to each element of {\tt A}, where {\tt misc} can be: \\
 & {\tt pow}, {\tt exp}, {\tt log}, {\tt log10}, {\tt sqrt}, {\tt square}, $\ldots$ \\
{\tt trig(A)} & apply a trigonometric function to each element of {\tt A}, where {\tt trig} can be: \\
 & {\tt cos}, {\tt sin}, {\tt tan}, {\tt acos}, {\tt asin}, {\tt atan}, $\ldots$ \\
\hline
\end{tabular}
\end{center}
\caption{Matrix valued functions (subset).}
\label{tab:func_matrices}
\end{table}

\clearpage

\section{Fundamental Differences from Armadillo}
\label{sec:arma_comparison}

Although Bandicoot aims to be API-compatible with Armadillo wherever possible,
major differences in the computational designs of CPUs and GPUs mean that
Bandicoot and Armadillo have some fundamental differences, as elucidated below.

The first and most important difference is
the location where data is stored:
in Armadillo, elements of a matrix are stored in CPU (host) memory;
that is, system RAM.
In Bandicoot, elements of a matrix are instead stored in device (GPU) memory.
Because Bandicoot users are writing control code that runs on the host,
this means that accessing the individual elements of a matrix requires a transfer from GPU memory to CPU memory.
For a single element, this can be very expensive!

To demonstrate the difference by counterexample,
consider the following example code:
{\footnotesize
\begin{minted}{c++}
    fmat A(1000, 1000, fill::randu);

    // count the number of elements greater than 0.3 (inefficient version)
    size_t count = 0;
    for (size_t i = 0; i < A.n_elem; ++i)
      {
      if (A[i] > 0.3) // WARNING: in Bandicoot, this causes a time-consuming CPU/GPU transfer!
        ++count;
      }
\end{minted}
}

When {\tt A} is an Armadillo matrix,
the matrix's elements are already in system memory and no transfer is necessary.
However, when {\tt A} is a Bandicoot matrix,
a transfer from GPU memory is performed.
On the same system used for benchmarks in Section~\ref{sec:speed},
this particular code snippet takes $0.0004$s with Armadillo
and $13$s with Bandicoot's CUDA backend!

For this reason, {\bf C++ iterators over Bandicoot matrices are not provided}.
This is an intentional design decision to prevent users from accidentally writing slow code.
Instead of iterating individually over elements in a Bandicoot matrix,
users should strive to use linear algebra expressions instead.
The code above can be more efficiently expressed as the equivalent operation:

{\footnotesize
\begin{minted}{c++}
    fmat A(1000, 1000, fill::randu);

    // count the number of elements greater than 0.3 (efficient version)
    size_t count = size(find(A > 0.3)).n_rows;
\end{minted}
}

Compared to the previous program,
this rewrite is more efficient (and more readable),
with speed comparable to Armadillo.

The second fundamental difference between Bandicoot and Armadillo
is related to the amount of parallelism that GPUs provide.
The main avenue to speedup with a GPU is provided by the extreme level of parallelism (SIMT).
For instance, Nvidia's recent GeForce RTX~4090 GPU contains 16,384 cores,
whereas recent AMD Epyc (x86-64 compatible) CPUs contain up to 128 cores
(i.e. about 2~orders of magnitude difference).
The clear implication of this is that keeping all of the cores in the GPU busy
is of the utmost importance to get speedup.

For this reason, Bandicoot is ill-suited to calculations on small matrices.
For instance, performing an operation like {\tt A += 3}
where {\tt A} is a $10 \times 10$ matrix
only makes use of at most 100 GPU cores.
Thus, if a program makes use of many small matrices,
conversion to Bandicoot may not give any speedup.
For this reason, fixed-size matrices (e.g. {\tt arma::mat::fixed<>}) do not provide tangible benefit
and are not currently implemented as a feature in Bandicoot.
As observed in our empirical evaluations shown in Section~\ref{sec:speed},
Bandicoot typically becomes noticeably faster than the CPU for matrices
sized about $1000 \times 1000$  
or larger;
the specific trade-off point will depend on the workload, CPU, and GPU,
so this `rule of thumb' above may not always apply.

The third fundamental difference between Bandicoot and Armadillo
is that GPUs tend to provide higher performance for lower-precision numeric types~\cite{haidar2018harnessing}.
Some older GPUs, in fact, do not have support for 64-bit floating point types~\cite{whitehead2011precision}.
For this reason, when using Bandicoot,
it is generally recommended to use 32-bit floating point numbers (e.g. {\tt float}) instead of 64-bit (e.g. {\tt double});
that is, the use of {\tt coot::fmat} matrix type is generally recommended instead of {\tt coot::mat} matrix type.
If {\tt coot::mat} is used but Bandicoot detects that the GPU does not support 64-bit precision,
a {\tt std::runtime\_error} exception will be thrown.

The last major fundamental difference between Bandicoot and Armadillo
is the fact that individual GPU kernels must be compiled at runtime for each specific device.
With Armadillo, all operations can be compiled entirely at compile time
(e.g. in the {\tt g++} compilation call).
In contrast, when any Bandicoot program is run,
the first step is to compile all GPU kernels for the device that is found.
Bandicoot caches the compiled kernels (by default in {\tt {\textasciitilde}/.bandicoot} on Linux and macOS systems),
so on most systems this startup step is a one-time cost that typically takes
between 3 and 5 minutes.
At the current time,
compilation of custom kernels for operations such as Armadillo's {\tt transform()} function
are not available,
but future releases aim to include this support (see Section~\ref{sec:roadmap}).

For situations where mixed use of Armadillo and Bandicoot is appropriate,
the {\tt conv\_to()} function can be used to convert back and forth, as below:

{\footnotesize
\begin{minted}{c++}
    using namespace coot;

    fmat X_gpu(1000, 1000, fill::randu);

    // transfer matrix from GPU to CPU
    arma::fmat X_cpu = conv_to<arma::fmat>::from(X_gpu);

    // transfer back from CPU to GPU
    X_gpu = conv_to<fmat>::from(X_cpu);
\end{minted}
}

\section{Linking Bandicoot Programs}
\label{sec:linking}

Like Armadillo, Bandicoot provides a {\it wrapper library} to ease the linking process.
Linking to GPU libraries (especially in the CUDA framework) can be cumbersome,
with Nvidia even going so far as providing the {\tt nvcc} compiler wrapper to simplify the process\footnote{\url{https://docs.nvidia.com/cuda/cuda-compiler-driver-nvcc/index.html}}.
Bandicoot's library obviates the need for this complexity,
and a typical program using Bandicoot can be compiled with a command as simple as

{\footnotesize
\begin{minted}{sh}
    g++ prog.cpp -o prog -O2 -lbandicoot
\end{minted}
}

This links against the Bandicoot wrapper library
({\tt libbandicoot.so} on Linux and {\tt libbandicoot.dylib} on macOS),
which is itself dynamically linked against all of Bandicoot's dependencies.
All of the functions in the Bandicoot wrapper library are simply
wrapper functions that in turn call the external library functions Bandicoot depends on.
Therefore, via transitive linking, a user can link against only the Bandicoot wrapper library.

An important note is that while Bandicoot and Armadillo are interoperable
and meant to be compatible,
Bandicoot does not depend on Armadillo.
Thus, there is only need to link against Armadillo ({\tt -larmadillo})
if Armadillo is explicitly being included and used in the program.

In some situations, it is preferable to not link against the wrapper library and
instead link directly against Bandicoot's dependencies.
This situation could occur if a user wants to use Bandicoot without installing the library,
or wants to use Bandicoot as a standalone header-only library.
If this is desired, the {\tt COOT\_DONT\_USE\_WRAPPER} macro must be set
(e.g. {\tt \#define COOT\_DONT\_USE\_WRAPPER}),
Then, the libraries that must be linked against are detailed below.

\begin{itemize}
  \itemsep -2pt
  \item Regardless of which backend is enabled, the following libraries must be linked against.
    \subitem {\tt -lblas} (CPU BLAS support; can use {\tt -lopenblas} instead)
    \subitem {\tt -llapack} (CPU LAPACK support; can use {\tt -lopenblas})

  \item If the OpenCL backend is enabled (e.g. {\tt COOT\_USE\_OPENCL} is defined), the following libraries are required.
    \subitem {\tt -lOpenCL} (core OpenCL support)
    \subitem {\tt -lclBLAS} (clBLAS for BLAS operations)

  \item If the CUDA backend is enabled (e.g. {\tt COOT\_USE\_CUDA} is defined), the following libraries are required.
    \subitem {\tt -lcuda} (core CUDA support)
    \subitem {\tt -lcudart} (CUDA runtime library)
    \subitem {\tt -lnvrtc} (runtime compilation of CUDA kernels)
    \subitem {\tt -lcublas} (cuBLAS for BLAS operations)
    \subitem {\tt -lcusolver} (cuSolverDn for decompositions and factorisations)
    \subitem {\tt -lcurand} (cuRand for random number generation)
\end{itemize}

\section{Empirical Speed Comparison}
\label{sec:speed}

We performed several tests in order to demonstrate the speedups that can be seen when using Bandicoot.
The speedups that are observed are not simply due to underlying speedups from
the use of GPUs instead of CPUs:
internally, Bandicoot uses delayed evaluation~\cite{Watt_2006},
where C++ template meta-programming techniques apply optimisations at compile time~\cite{Vandevoorde_2017}.
Where applicable,
the order of operations is optimised,
while redundant or unnecessary operations are removed entirely,
and inefficiently-expressed operations are rewritten efficiently.
This technique is also used in Armadillo and has been shown to
provide significant speedups over alternate linear algebra libraries~\cite{Psarras_2022,sanderson2017armadillo}.

Our timing comparisons focus on the speedups Bandicoot can provide over Armadillo
due to the use of GPUs instead of CPUs.
These comparisons also shed light on the situations where Bandicoot is more appropriate than Armadillo.
All comparisons were done on a system with an Intel Core i9-10920X (3.5 GHz, 12
cores, 19.25MB cache),
an Nvidia RTX~2080Ti GPU (1350 MHz, 4352 cores, 11GB RAM) with Nvidia driver version 525.105.17,
and 128GB system RAM,
with Linux kernel 6.1.0 and GCC 12.2.0.
For completeness, comparisons were performed using both 32-bit floating-point precision ({\tt float})
and 64-bit floating-point precision ({\tt double}).
(See Section~\ref{sec:arma_comparison} for more details on precision on GPUs.)

For these comparisons,
we use Bandicoot's OpenCL backend,
Bandicoot's CUDA backend,
and Armadillo on the CPU.

\newpage

{\bf Task 1}. Accumulate elements of a vector.  The inner loop of
the code used for timing is shown below.

{\footnotesize
\begin{minted}{c++}
    // N is specified externally
    // when using Armadillo, 'using namespace arma' is used
    // when using Bandicoot, 'using namespace coot' is used
    fvec f(N);
    f.randu();
    coot_synchronise(); // applicable only to Bandicoot

    wall_clock c;

    c.tic();
    float result = accu(c);
    coot_synchronise(); // applicable only to Bandicoot
    const double time = c.toc();
\end{minted}
}

\begin{figure}[h!]
\centering
\begin{subfigure}{.49\textwidth}
  \centering
  \includegraphics[width=0.95\textwidth]{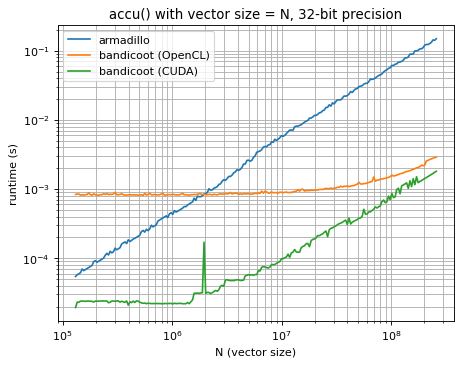}
  \caption{{\tt accu()} performance for {\tt float}s.}
  \label{fig:accu_float}
\end{subfigure}
\begin{subfigure}{.49\textwidth}
  \centering
  \includegraphics[width=0.95\textwidth]{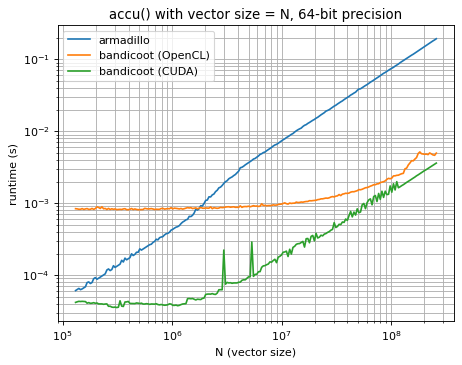}
  \caption{{\tt accu()} performance for {\tt double}s.}
  \label{fig:accu_double}
\end{subfigure}
\caption{{\tt accu()} performance for Bandicoot and Armadillo as a function of vector size.}
\label{fig:accu}
\end{figure}

Figure~\ref{fig:accu} shows results for the {\tt accu()} task.
The observations about matrix size from Section~\ref{sec:arma_comparison} are validated:
Bandicoot is faster only for larger vector sizes.
The CUDA backend is more effective than the OpenCL backend,
which is not particularly surprising on an Nvidia device.
For large enough vector sizes, Bandicoot can outperform Armadillo by roughly two orders of magnitude.

\newpage

{\bf Task 2.} Scalar vector multiply-add ({\it axpy}).  The inner loop of the code used for timing is shown below.

{\footnotesize
\begin{minted}{c++}
    // N is specified externally
    // when using Armadillo, 'using namespace arma' is used
    // when using Bandicoot, 'using namespace coot' is used
    fvec A(N, fill::randu);
    fvec B(N, fill::randu);
    coot_synchronise(); // applicable only to Bandicoot

    wall_clock c;

    c.tic();
    B += 3 * A;
    coot_synchronise(); // applicable only to Bandicoot
    const double time = c.toc();
\end{minted}
}

\begin{figure}[h!]
\centering
\begin{subfigure}{.49\textwidth}
  \centering
  \includegraphics[width=0.95\textwidth]{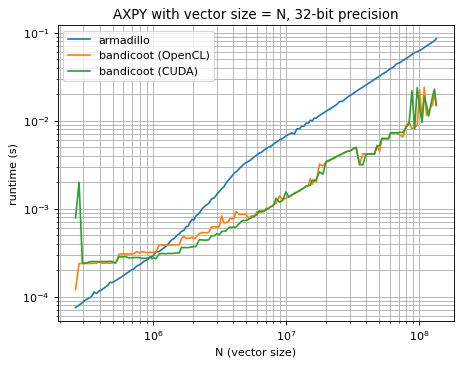}
  \caption{{\it axpy} performance for {\tt float}s.}
  \label{fig:axpy_float}
\end{subfigure}
\begin{subfigure}{.49\textwidth}
  \centering
  \includegraphics[width=0.95\textwidth]{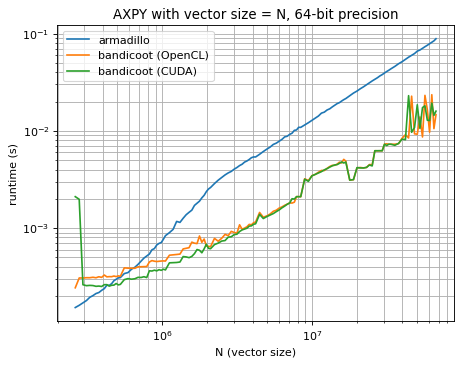}
  \caption{{\it axpy} performance for {\tt double}s.}
  \label{fig:axpy_double}
\end{subfigure}
\caption{{\it axpy} performance for Bandicoot and Armadillo as a function of vector size.}
\label{fig:axpy}
\end{figure}

The {\it axpy} results in Figure~\ref{fig:axpy} match the observations from the {\tt accu()} experiment:
Bandicoot is only faster for larger vector sizes.

\newpage

{\bf Task 3.} Matrix multiplication.  The inner loop of the code used for timing
is shown below.

{\footnotesize
\begin{minted}{c++}
// N is specified externally
// when using Armadillo, 'using namespace arma' is used
// when using Bandicoot, 'using namespace coot' is used
fmat A(N, N, fill::randu);
fmat B(N, N, fill::randu);
coot_synchronise(); // applicable only to Bandicoot

wall_clock c;

c.tic();
fmat C = A * B;
coot_synchronise(); // applicable only to Bandicoot
const double time = c.toc();
\end{minted}
}

\begin{figure}[h!]
\centering
\begin{subfigure}{.49\textwidth}
  \centering
  \includegraphics[width=0.95\textwidth]{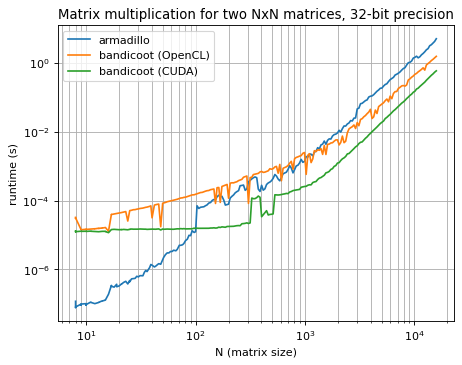}
  \caption{Matrix multiplication performance for {\tt float}s.}
  \label{fig:matmul_float}
\end{subfigure}
\begin{subfigure}{.49\textwidth}
  \centering
  \includegraphics[width=0.95\textwidth]{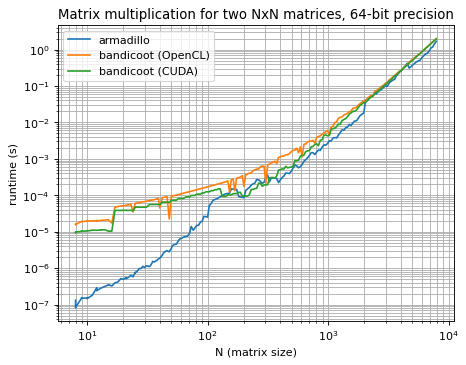}
  \caption{Matrix multiplication performance for {\tt double}s.}
  \label{fig:matmul_double}
\end{subfigure}
\caption{Matrix multiplication performance for Bandicoot and Armadillo with square matrices as a function of matrix size.}
\label{fig:matmul}
\end{figure}

The results in Figure~\ref{fig:matmul} highlight some additional differences between CPUs and GPUs:
the performance differences for {\tt float} and {\tt double} element types,
discussed in Section~\ref{sec:arma_comparison},
are clear in this benchmark.
32-bit performance significantly exceeds 64-bit performance,
and for 64-bit types, on this system, arguably no speedup is seen.
Note that because Bandicoot and Armadillo wrap other libraries where possible,
this comparison is really a comparison between clBLAS~\cite{clblas}, cuBLAS, and OpenBLAS~\cite{OpenBLAS_2023,wang2013augem}.

\newpage

{\bf Task 4.} LU decomposition.  The inner loop of the code used
for timing is given below.

{\footnotesize
\begin{minted}{c++}
// N is specified externally
// when using Armadillo, 'using namespace arma' is used
// when using Bandicoot, 'using namespace coot' is used
fmat A(N, N, fill::randu);
coot_synchronise(); // applicable only to Bandicoot

wall_clock c;

c.tic();
fmat L, U;
lu(L, U, A);
coot_synchronise(); // applicable only to Bandicoot
const double time = c.toc();
\end{minted}
}

\begin{figure}[h!]
\centering
\begin{subfigure}{.49\textwidth}
  \centering
  \includegraphics[width=0.95\textwidth]{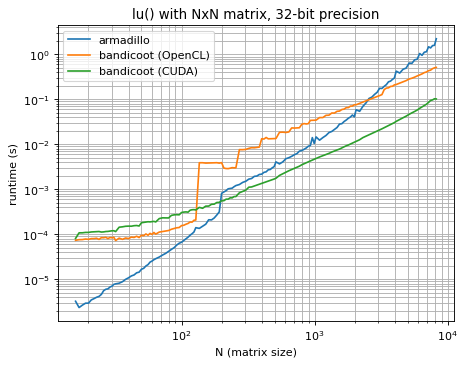}
  \caption{{\tt lu()} performance for {\tt float}s.}
  \label{fig:lu_float}
\end{subfigure}
\begin{subfigure}{.49\textwidth}
  \centering
  \includegraphics[width=0.95\textwidth]{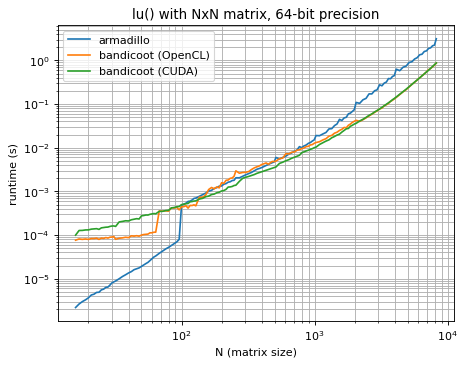}
  \caption{{\tt lu()} performance for {\tt double}s.}
  \label{fig:lu_double}
\end{subfigure}
\caption{LU decomposition performance for Bandicoot and Armadillo with square matrices as a function of matrix size.}
\label{fig:lu}
\end{figure}

Due to the more complex nature of decompositions,
they are not as easily parallelized onto the GPU as operations like {\tt accu()} and {\it axpy},
and Figure~\ref{fig:lu} shows that speedups may be more limited in this case.
Nonetheless, Bandicoot is still able to provide up to a 10x speedup for {\tt float} matrices,
and up to a 5x speedup for {\tt double} matrices,
for matrices that are sufficiently large.

\newpage

\section{Roadmap}
\label{sec:roadmap}

So far as is possible,
the primary medium-term goal for Bandicoot is to be fully API-compatible with Armadillo,
to ease conversion of Armadillo programs.
However, Bandicoot opens numerous possibilities to go beyond Armadillo.
Some of the roadmap goals of the development team are listed and detailed below,
in approximate priority order.
(Note that priorities may change over time;
the development team does not intend to address this list sequentially.)

\begin{itemize}
  \item {\bf Additional decompositions}.  Bandicoot does not currently implement
the {\tt qr()}, {\tt eig\_gen()}, {\tt qz()}, or {\tt schur()} decompositions.
Bandicoot also lacks some of the more complex options of Armadillo's {\tt
solve()}~\cite{sanderson2020adaptive}.

  \item \textbf{Support for {\tt Cube}}.  Bandicoot does not currently implement the
3-dimensional matrix class {\tt Cube}.
This class is especially useful for higher-order operations that operate on collections of matrices.

  \item {\bf Complex number support}.  Bandicoot does not currently allow complex element types for matrices, e.g.,
{\tt Mat<~std::complex<float>~>}.
It is a near-term goal to support this,
as this is important and highly-used functionality in Armadillo.

  \item {\bf Use of CLBlast instead of clBLAS}.  The clBLAS project~\cite{clblas}
has not been updated in years and there are some bugs that are likely to remain
unfixed.
The newer CLBlast project~\cite{nugteren2018clblast} provides an appealing alternative.
Like Armadillo's support for any BLAS/LAPACK replacement,
the goal for Bandicoot is to allow the use of either library with the OpenCL backend.

  \item {\bf Additional backends}.  The balkanization of the GPU space is a real
concern, as OpenCL is effectively being replaced by proprietary frameworks from multiple vendors.
Although OpenCL is likely to work as a backend for many years,
better performance will be obtained with vendor-specific backends
due to the amount of effort those vendors put into optimising their specific
backends.
A few examples of backends that would be useful to add are listed below.

\begin{itemize}
\item {\it Metal}. The Apple Metal framework\footnote{\url{https://developer.apple.com/metal/}}
is Apple's new accelerator framework for hardware acceleration.
\item {\it ROCm}. AMD has an accelerator platform tuned for its GPUs called
ROCm\footnote{\url{https://www.amd.com/en/graphics/servers-solutions-rocm}}.
\item {\it oneAPI}. Intel has its own competing framework called
oneAPI\footnote{\url{https://www.intel.com/content/www/us/en/developer/tools/oneapi/overview.html}}.
\item {\it SYCL}. The SYCL framework\footnote{\url{https://www.khronos.org/sycl/}}
is not vendor-specific, and could be considered a spiritual successor to OpenCL.
\end{itemize}

  \item {\bf Low-precision type support}.
Especially in the field of machine learning,
low-precision floating-point types are often used to increase computational
performance and/or energy efficiency~\cite{hubara2017quantized,tagliavini2018transprecision}.
Types such as `brain floating point' have also been gaining widespread adoption~\cite{Burgess_2019}.
As CUDA supports precisions down to 8-bit ({\tt fp8}),
and OpenCL extensions support half-precision floating-point ({\tt fp16}),
it is an important development goal to allow Bandicoot to seamlessly support these types
whenever they are available in hardware.

  \item {\bf Integration into higher-level libraries}.
With the first release of Bandicoot now complete,
work is underway to adapt higher-level libraries that use Armadillo
to also be capable of using Bandicoot.
Two important examples are the
\textit{ensmallen} C++ numerical optimisation library~\cite{curtin2021ensmallen}
and the
\textit{mlpack} C++ machine learning library~\cite{curtin2023mlpack}.
Other possibilities being considered for development also include
an Rcpp binding~\cite{rcpp} for Bandicoot to allow users of R to GPU-accelerate their RcppArmadillo code~\cite{eddelbuettel2014rcpparmadillo}.
\end{itemize}

More development details and plans can be found on the Gitlab development website,
found at \url{https://gitlab.com/conradsnicta/bandicoot-code/}.

\vfill

\section{Internal Design}
\label{sec:internals}

Internally, Bandicoot is built on a template meta-programming framework similar to Armadillo's template meta-programming framework.
The idea of this framework is to provide significant compile-time optimisations of mathematical expressions given in user code.
This is accomplished by collecting (at compile time) the structure of a linear
algebra expression as a type,
and then, when possible,
using techniques such as SFINAE~\cite{Vandevoorde_2017} and template specialisation
to restructure the expression into an equivalent but more efficient form.
Importantly, to keep Bandicoot user-friendly,
this entire template meta-programming framework is internal to the library,
and therefore invisible to the user.

In this section, we give an overview of that framework and
show some simple examples of how optimisations can be applied in this manner.
We also discuss differences between the internal designs of Bandicoot and Armadillo.
The discussion is aimed to be useful to those who are curious about the internal details,
or those who are interested in contributing to Bandicoot development.
Knowledge of C++ and template meta-programming concepts is helpful.
This section is not necessary reading for Bandicoot users:
the documentation in earlier sections of this report and on the Bandicoot website is sufficient for that use case.

The base type in Bandicoot,
which is the base type that all matrix and vector objects as well as all linear algebra expressions inherit from,
is the template type {\tt Base<eT, T1>} where
{\tt eT} refers to the {\it element type} and takes values {\tt float}, {\tt
double}, {\tt int}, etc.,
and {\tt T1} refers to the actual type of the Bandicoot linear algebra expression itself.
This is a form of static polymorphism
(that is, polymorphism without runtime virtual-table lookups)
modeled after the ``Curiously Recurring Template Pattern'' (CRTP)~\cite{alexandrescu2001modern, coplien1996curiously}.

User-facing classes include the {\tt Mat<eT>}, {\tt Col<eT>}, and {\tt Row<eT>} classes,
which represent matrices, column vectors, and row vectors, respectively.
A number of predefined convenience {\tt typedef}s are given,
allowing users to use, e.g., the more concise {\tt fmat} type instead of {\tt Mat<float>}.

Expressions in Bandicoot are built up as C++ types using the {\tt Op<eT, T1, op\_type>},
{\tt Glue<eT, T1, T2, glue\_type>},
{\tt mtOp<out\_eT, T1, mtop\_type>},
{\tt eOp<eT, T1, eop\_type>},
and {\tt eGlue<eT, T1, T2, eglue\_type>} template classes.
Each of these classes represents a specific type of linear algebra operation:

\begin{itemize}
  \itemsep -2pt
  \item {\small\tt Op<T1, op\_type>}: unary operation of type {\tt op\_type} that will operate on {\tt T1}
  \item {\small\tt mtOp<out\_eT, T1, mtop\_type>}: a multi-type operation of type {\tt mtop\_type} on {\tt T1} that produces an output that has different element type {\tt out\_eT}
  \item {\small\tt eOp<T1, eop\_type>}: an element-wise operation of type {\tt eop\_type} on {\tt T1}
  \item {\small\tt Glue<T1, T2, glue\_type>}: binary ``glue'' operation of type {\tt glue\_type} that will operate on {\tt T1} and {\tt T2}
  \item {\small\tt eGlue<T1, T2, eglue\_type>}: an element-wise ``glue'' operation of type {\tt eglue\_type} on two operands of types {\tt T1} and {\tt T2}
\end{itemize}

Here, the operation types ({\tt op\_type}, {\tt glue\_type}, etc.)
are placeholder types that represent the operation to be performed:
for instance, Bandicoot contains types such as {\tt op\_sum}, {\tt glue\_times}, {\tt mtop\_conv\_to}, {\tt op\_vectorise}, {\tt op\_repmat}, and so forth.

Expressions are assembled as users call functions.
Bandicoot's functions do not return types such as {\tt Mat<eT>},
but instead return any of the expression types listed above,
which capture (as a type) the expression to be performed.
As an example, the function signature for Bandicoot's {\tt trans()} function is as follows:

{\footnotesize
\begin{minted}{c++}
    // Note: the actual version in the library may differ slightly.
    template<typename eT, typename T1>
    inline
    const Op<T1, op_htrans>
    trans(const Base<eT, T1>& X);
\end{minted}
}

The important observations here are that {\tt trans()} does not require a matrix or vector object as input---instead, it can receive any Bandicoot expression;
and, instead of returning a {\tt Mat<eT>} that holds the transposed result,
the function returns an \mbox{\small\tt Op<T1, op\_htrans>} indicating that
computing the value of the expression will require taking the transpose of {\tt T1}.

Before considering how such expressions are evaluated,
we can easily demonstrate how some compile-time optimisations can be performed.
Observe the following additional overload of the {\tt trans()} function that Bandicoot includes:

{\footnotesize
\begin{minted}{c++}
    // Note: the actual version in the library may differ slightly.
    template<typename eT, typename T1>
    inline
    typename
    enable_if2
      <
      is_coot_type<T1>::value,
      const Op<T1, op_diagmat>
      >::result
    trans(const Op<T1, op_diagmat>& X);
\end{minted}
}

This more complicated overload is for situations where a user passes a diagonal matrix
(signified as an {\tt op\_diagmat}; this is produced by e.g. the {\tt diagmat()} function).
Because the transpose of a diagonal matrix is equivalent to the diagonal matrix itself,
we do not even need to add an {\tt op\_htrans} to the expression type,
and so we can simply return the given {\tt Op<T1, op\_diagmat>} completely unchanged.
Note that the {\tt enable\_if2<>} is an SFINAE restrictor on the return type~\cite{Vandevoorde_2017}
that will not allow the function to be selected unless the given {\tt T1} is a Bandicoot type
(e.g. it must inherit from {\tt Base<eT, T1>} for some {\tt eT}).

Bandicoot contains multiple optimisations of this form,
where user-level functions will greedily apply optimisations
during the construction of the full expression type.

As an example of how compound expressions are represented internally, 
let us consider the following expression:

{\footnotesize
\begin{minted}{c++}
    A * trans(diagmat(V)) + 4 * B
\end{minted}
}

where {\tt A} and {\tt B} have type {\tt mat}, and {\tt V} has type {\tt vec}.
Abusing notation to replace individual operations with their resulting type,
the type expression formed by the compiler for the input above is

{\footnotesize
{\tt
\ \ \ \ A * trans(diagmat(V)) + 4 * B

\vspace*{-1em}
\ \ \ \ A * trans({\color{red}{Op<vec, op\_diagmat>}}) + 4 * B

\vspace*{-1em}
\ \ \ \ A * {\color{red}{Op<vec, op\_diagmat>}} + 4 * B

\vspace*{-1em}
\ \ \ \ {\color{red}{Glue<mat, Op<vec, op\_diagmat>, glue\_times>}} + 4 * B

\vspace*{-1em}
\ \ \ \ {\color{red}{Glue<mat, Op<vec, op\_diagmat>, glue\_times>}} + {\color{red}{eOp<mat, eop\_scalar\_times>}}

\vspace*{-1em}
\ \ \ \ {\color{red}{eGlue<Glue<mat, Op<vec, op\_diagmat>, glue\_times>, eOp<mat, eop\_scalar\_times>, eglue\_plus>}}
}
}

While this type is quite intimidating in its textual representation
(especially when encountered as an error message during Bandicoot development),
it is a complete description of the operation to be performed.
Further, with most compilers, these types are ephemeral and can be completely optimised out of the program.
Each operation type ({\tt Op}, {\tt Glue}, etc.) simply holds a reference to its operands,
and possibly one or two primitives ({\tt int}, etc.) of metadata,
and thus may be easily optimised out of the program entirely.

Next, let us consider how expressions like this are evaluated and the results stored in specified output matrices.
The expression above may appear in a program like this:

{\footnotesize
\begin{minted}{c++}
    mat C = A * trans(diagmat(V)) + 4 * B;
\end{minted}
}

The evaluation of Bandicoot expressions is delayed until an assignment operator is encountered.
In this case, the function that will actually cause the expression to be evaluated is
{\tt Mat<double>::operator=(const Base<eT, T1>\&)}
where {\tt T1} will be the full type of the expression, given above.

The majority of the work that {\tt operator=(const Base<eT, T1>\&)} performs is handled by the
{\tt unwrap<T1>} meta-programming struct.
{\tt unwrap} uses partial template specialisations, each dependent on the outer type of the given expression.
The simplest {\tt unwrap} specialisation is for a {\tt Mat} object:

\newpage 
{\footnotesize
\begin{minted}{c++}
    // Note: this is a conceptual simplification
    template<typename eT>
    struct unwrap< Mat<eT> >
      {
      const Mat<eT>& M;

      unwrap(const Mat<eT>& in) : M(in) { }
      };
\end{minted}
}

When {\tt operator=($\ldots$)} is called on a {\tt Mat<eT>} argument,
{\tt unwrap} only stores a reference to the matrix via the {\tt .M} member,
followed by {\tt operator=($\ldots$)} accessing this result.

More complicated unwrapping behavior is performed for expressions.
For instance, the {\tt unwrap} struct specialised for a given type named as {\tt Op<T1, op\_type>}
will call the static function {\tt op\_type::apply(Mat<eT>\&, Op<T1, op\_type>\&)},
which is expected to compute the result of the operation on the input {\tt T1}
and store it in the given output matrix;
for example, {\tt op\_diagmat} will generate a diagonal matrix from a given vector expression.
The specific {\tt apply()} function implemented for {\tt op\_type}
will then also use the {\tt unwrap} struct on the given {\tt T1} type,
which in turn may result in follow-on calls to other {\tt apply()} functions
which use partially specialised {\tt unwrap} structs.
Additional optimisations via partial specialisations of the arguments of the {\tt apply()} function may also be applied.

One optimisation that is unique to Bandicoot is its ability to fuse type conversions into operations.
Consider the following example code:

{\footnotesize
\begin{minted}{c++}
  fmat B = conv_to<fmat>::from(resize(A, 5, 6));
\end{minted}
}

Assuming that {\tt A} has type {\tt imat} (i.e. integer matrix),
the type of this expression is

{\footnotesize
\begin{minted}{c++}
  mtOp<float, Op<Mat<int>, op_resize>, mtop_conv_to>
\end{minted}
}

As with all other operations, {\tt unwrap} will first call the function
{\tt mtop\_conv\_to::apply()} with {\tt B} as the output matrix.
However, instead of first evaluating the inner operation and then converting the result,
Bandicoot will {\it fuse} the inner operation with the type conversion.
This is accomplished at the top level by allowing each {\tt apply()} function
to output into a matrix of a different element type.
So, in this example, the function
{\tt op\_resize::apply(Mat<float>\&, Op<Mat<int>, op\_resize>\&)},
which has different output and input element types,
will be called.

Subsequently, {\tt op\_resize::apply()},
when actually performing the resize operation (merely a copy of the elements),
will call a ``two-way'' GPU kernel,
where the input element type is allowed to differ from the output element type,
and the conversion between input and output types is performed as part of the computation.
The use of this two-way kernel thus prevents an additional copy of elements after the original resizing operation.
In this example, fusing the conversion cuts the operation time in half.

Many operations in Bandicoot are implemented with support for fused conversions.
The motivation for this is to reduce the number of separate GPU operations
necessary to evaluate an expression.
Whereas Armadillo is able to generate code specific to each individual expression,
Bandicoot is not (currently) able to generate GPU code specific to each individual expression;
therefore, reducing the number of GPU operations as much as possible is a priority.

Typically,
after template meta-programming optimisations and fused conversions are applied to the expression in the manners described above,
an individual operation's {\tt apply()} function will call at least one GPU kernel using
either the OpenCL or CUDA backend.
This has conceptual similarity to Armadillo, whose ``backend''
includes direct calls to LAPACK and BLAS functions.

Since Bandicoot supports multiple backends,
which can be configured at runtime,
a singleton object exists as an interface layer
between the {\tt apply()} functions and the GPU.
The object is called {\tt coot\_rt\_t} and is obtained with {\tt coot\_rt\_t::get\_rt()}.
{\tt coot\_rt\_t} provides numerous {\tt static} functions that take
direct GPU memory as input.
Thus, a typical {\tt apply()} function implementation may contain calls such as,
e.g., {\tt coot\_rt\_t::acquire\_memory<eT>(n\_elem)} to allocate memory,
or {\tt coot\_rt\_t::accu(X.get\_dev\_mem(), X.n\_elem)}
to accumulate all the elements in a given object {\tt X}.

The {\tt coot\_rt\_t} object also serves as an implementational firewall
between high-level Bandicoot objects and the GPUs.
All CUDA-specific code is contained entirely within the
{\tt coot\_rt\_t::get\_rt().cuda\_rt} object,
and all OpenCL-specific code,
including calls to supporting libraries such as clBLAS,
is contained entirely within the
{\tt coot\_rt\_t::get\_rt().cl\_rt} object.
This separation makes it easy to enable or disable a backend at compile time,
as well as to add an entirely new backend (as alluded to in Section~\ref{sec:roadmap}).

This architectural overview provides the basics of the internal structure of Bandicoot.
We note that it is a cursory look at the details of Bandicoot's implementation,
and hence it is intended as a starting point for further exploration of the codebase.

\section{Conclusion}
\label{sec:conclusion}

In this report we provided an overview of the Bandicoot C++ GPU linear algebra library.
Bandicoot aims to provide an API that is compatible with Armadillo's API,
allowing users to take advantage of GPU-accelerated computation for their existing codebases
without significant changes.
Empirical evaluations show that Bandicoot can provide significant speedups over CPU-only computation,
depending on the nature of the workload and the size of the data.
Bandicoot can use either an OpenCL or CUDA backend.

Bandicoot's open-source nature enables extensions and customisation,
and we hope that the community will find it useful.
Contributions (especially towards goals in the given roadmap of Section~\ref{sec:roadmap}) are more than welcome
and can be made on the Bandicoot Gitlab code repository:
\url{https://gitlab.com/conradsnicta/bandicoot-code/}.
More details about the library,
including frequently asked questions
and comprehensive documentation of user-accessible functions
can be found on the publicly accessible website at \url{https://coot.sourceforge.io}.


\small

\bibliographystyle{ieee}
\bibliography{refs}

\end{document}